\documentclass[pra,twocolumn,showpacs]{revtex4-1}
\usepackage{amsmath,amscd,amssymb,color}
\usepackage{graphicx,amsfonts,epsf}
\usepackage{epstopdf}
\usepackage{float}
\usepackage{enumerate}
\usepackage{relsize}
\usepackage{mathtools}
\usepackage{physics}
\usepackage{array,multirow}
\usepackage{array}
\usepackage{multirow}

\usepackage{xcolor}
\definecolor{maroon}{RGB}{100,20,20}
\definecolor{dblue}{RGB}{20,20,100}
\usepackage{ulem}
\usepackage{multirow}
\begin{document}
\title{Using linear and nonlinear entanglement witnesses  
to generate and detect bound entangled states
on an IBM quantum processor}
\author{Vaishali Gulati}
\email{vaishali@iisermohali.ac.in}
\affiliation{Department of Physical Sciences, Indian Institute of Science Education \& Research Mohali,
Sector 81 SAS Nagar, Manauli PO 140306 Punjab India}
\author{Gayatri Singh}
\email{ph20015@iisermohali.ac.in}
\affiliation{Department of Physical Sciences, Indian Institute of Science Education \& Research Mohali,
Sector 81 SAS Nagar, Manauli PO 140306 Punjab India}
\author{Kavita Dorai}
\email{kavita@iisermohali.ac.in}
\affiliation{Department of Physical Sciences, Indian Institute of Science Education \& Research Mohali,
Sector 81 SAS Nagar, Manauli PO 140306 Punjab India}
\begin{abstract}
We investigate bound entanglement in three-qubit mixed states which are
diagonal in the Greenberger-Horne-Zeilinger (GHZ) basis.  Entanglement
in these states is detected using entanglement witnesses and the
analysis focuses on states exhibiting positive partial transpose (PPT).
We then compare the detection capabilities of optimal linear and
nonlinear entanglement witnesses. In theory, both linear and nonlinear
witnesses produce non-negative values for separable states and negative
values for some entangled GHZ diagonal states with PPT, indicating the
presence of entanglement. Our experimental results reveal that in cases
where linear entanglement witnesses fail to detect entanglement,
nonlinear witnesses are consistently able to identify its presence.
Optimal linear and nonlinear witnesses  were generated on an IBM
quantum computer and their performance was evaluated using two bound
entangled states (Kay and Kye states) from the literature, and randomly
generated entangled states in the GHZ diagonal form.  Additionally, we
propose a general quantum circuit for generating a three-qubit GHZ
diagonal mixed state using a six-qubit pure state on the IBM quantum
processor. We experimentally implemented the circuit to obtain
expectation values for three-qubit mixed states and compute the
corresponding entanglement witnesses.
\end{abstract}
\maketitle
\section{Introduction}
Entanglement is a core concept in quantum
theory~\cite{horodecki-rmp-2009,guhne-pr-2009,das-qibook-2016} which drives
quantum information processing and computing. Detecting
entanglement~\cite{altepeter-prl-2005,van-pra-2007} is challenging, especially
for larger systems where measurement scalability is unfavorable. Refined
detection methods such as the correlation matrix
criterion~\cite{vicente-qic-2007,vicente-jpa-2008}, the covariance matrix
criterion~\cite{guhne-prl-2007}, and entanglement witnesses
(EWs)~\cite{horodecki-pla-1996,terhal-pla-2000,brub-jomo-2002} have been used
for entanglement validation. The positive partial transpose (PPT) or the
Peres-Horodecki criterion~\cite{horodecki-pla-1996} fully addresses the
detection problem for 2$\otimes$2 and 2$\otimes$3 states, but difficulties
arise for systems with more than two qubits. PPT entangled states, also known
as bound entangled states, are positive under partial transposition and are
found in higher-dimensional systems. These states have interesting applications
in information concentration~\cite{murao-prl-2001}, secure key
distillation~\cite{horodecki-prl-2005,augusiak-pra-2006}, and have been used as
resources for certain zero-capacity quantum channels~\cite{smith-sciene-2008}.

A method to detect entanglement in higher-dimensional systems involves
entanglement witnesses (EWs)~\cite{terhal-pla-2000}.  In linear entanglement
witnesses, the expected value of a linear operator generates inequalities that
reveal entanglement. Violations of these inequalities indicate entanglement,
with positive values indicating separable states and negative values indicating
the presence of at least one entangled
state~\cite{horodecki-pla-2001,terhal-tcs-2002}. While conventional methods use
linear inequalities, more recent research has focused on nonlinear entanglement
witnesses~\cite{guhne-prl-2006,uffink-prl-2002,guhne-prl-2004,toth-pra-2005}.

GHZ diagonal states have been extensively studied in quantum information
processing, and feature a simple structure where only the diagonal and
anti-diagonal elements of the density matrix are
non-zero~\cite{dur-pra-2000(2),guhne-njp-2010,nagata-ijtp-2009,aolita-prl-2008}.
Due to decoherence and imperfections in state preparation, experimental
multipartite entangled states are typically mixed. They play a crucial role in
quantum information theory, including applications like quantum channel
capacity~\cite{kay-pra-2011,xychen-pra-2011}. The separability and
biseparability of GHZ diagonal states have also been
studied~\cite{kay-pra-2011,huber-prl-2010,christopher-prl-2012,guhne-pla-2011}.

The IBM quantum processor has been extensively used in quantum information
research ranging from simulating open quantum system
dynamics~\cite{garcia-npj-2020}, neutrino
oscillations~\cite{yeter-qip-2022,jha-epj-2022}, quantum effects of
gravity~\cite{manabputra-qip-2020,lyu-jocp-2021}, quantum chemistry
simulations~\cite{tilly-pra-2020,yeter-npj-2020}, Shor's factoring
algorithm~\cite{skosana-screp-2021} and quantum key
distribution~\cite{amico-pra-2019}. Recently an algorithm to measure
entanglement in a bipartite system was implemented on IBM's 127-qubit
system~\cite{karimi-ps-2024}.

In this work, we experimentally demonstrate the detection of entanglement in
mixed three-qubit states which are diagonal in the GHZ basis. We utilize both
linear and nonlinear entanglement witnesses, allowing us to effectively detect
PPT entangled states, also known as bound entangled states. This investigation
is carried out using the IBM Quantum Experience cloud service.  The states
under consideration in our study are mixed and are characterized by both
diagonal and non-diagonal entries in their density matrices. Due to their mixed
nature, these states cannot be directly prepared on an IBM quantum processor
using only three qubits.  We hence provide a general protocol where the
preparation involves three ancilla qubits.  The protocol begins with a
six-qubit register initialized in the state $\vert000000\rangle$. From this
initial state, we generate a six-qubit pure state such that the reduced density
operator of the first three qubits corresponds to the desired three-qubit GHZ
diagonal state. After experimentally generating these states, we demonstrate
that the generated states indeed carry bound entanglement. This entanglement
can be verified using an entanglement witness. Our study involves two
well-documented bound entangled states, namely the Kay
state~\cite{kay-pra-2011} and the Kye state~\cite{kye-jpa-2015}. Additionally,
we include a set of PPT entangled states which were randomly generated based on
the classifications provided in the Reference~\cite{Jafarizadeh-epjd-2008}.  We
find that the detection capability of nonlinear entanglement witnesses
surpasses that of their linear counterparts. This allows us to detect
entanglement in states where linear witnesses failed to do so.

This paper is organized as follows: Sec.~\ref{theory} provides an overview of
GHZ diagonal states with PPT, along with their linear and non-linear
entanglement witnesses. Several examples of PPT entangled states are discussed
in Sec.~\ref{kay-key}. The experimental details for the preparation and
implementation of entangled states and witness detection are described in
Sec.~\ref{exp1} and Sec.~\ref{exp2}, respectively. Finally, a few conclusions
are presented in Sec.~\ref{concl}.
\section{GHZ Diagonal States and Entanglement Witnesses}
\label{theory}
The density matrix of a three-qubit GHZ diagonal state is characterized by
non-zero diagonal and anti-diagonal entries, while all other entries are zero.
A three-qubit GHZ diagonal state can be written in the form
\begin{equation}
\rho =
\sum_{k=1}^8 p_k\vert
\phi_k 
\rangle \langle
\phi_k \vert \hspace{0.2cm}\text{ where, }
0 \leq  p_k \leq 1  \text{ }\&
\sum_{k=1}^8  p_k = 1
\label{eq1}
\end{equation}
where $p_k$ is the probability corresponding to $\ket{\phi_k}$. For three
qubits, the GHZ basis consists of eight vectors $\ket{\phi_k}
\\\vspace{-0.3cm}\\=\frac{1}{\sqrt{2}} (\ket{0x_ix_j} \pm
\ket{1\bar{x_i}\bar{x_j}})$, with $x_i, x_j \in (0, 1)$ and $x_i \neq
\bar{x}_i$.
In binary notation, $k-1 =0 x_i x_j$ for the
$`+$' states and $k-1 = 1\bar{x_i}\bar{x_j}$ for the $`-$'
states.

We can rewrite the GHZ diagonal
states given in Eq.~\ref{eq1}) 
using Pauli operators $\sigma_x, \sigma_y, \sigma_z$ 
and the $2 \cross 2$ Identity matrix $I$,  (omitting the tensor product sign) as
\begin{equation}
    \begin{split}
        \rho= &\frac{1}{8}
[III + r_1 \sigma_z\sigma_zI + r_2\sigma_zI\sigma_z + r_3 I\sigma_z\sigma_z + r_4\sigma_x\sigma_x\sigma_x \\
&+ r_5\sigma_x\sigma_y\sigma_y + r_6\sigma_y\sigma_x\sigma_y + r_7\sigma_y\sigma_y\sigma_x]
    \end{split}
    \label{eq2}
\end{equation}
This approach to detect entanglement in GHZ diagonal states is based on the PPT
criterion. If the state is entangled and its partial transposition with respect
to all bi-partitions is positive, it is referred to as a PPT entangled or a
bound entangled state.  A family of optimal linear entanglement witnesses (EWs)
which are obtained from the linear combinations of Pauli operators appearing in
the GHZ diagonal states (Eq. \ref{eq2}) have been introduced in
Ref.~\cite{Jafarizadeh-epjd-2008}, such that when applied on a pure product
state, the minimum value of the trace of EW results in 0. These EWs can detect
entanglement in GHZ diagonal density matrices with positive partial
transpositions.

The linear optimal EWs are expressed as
\begin{equation}
\fontsize{10pt}{10pt}\selectfont
    \mathcal{W}= III \pm \mathcal{O}_i + \cos\theta(\mathcal{O}_j\pm \mathcal{O}_k)+ \sin \theta(\mathcal{O}_l \pm \mathcal{O}_m)\label{eq3}
\end{equation}
where $i=1,2,3$ and $j,k,l,m=4,5,6,7$ with $j\neq k\neq l\neq m$. The
observables are defined as $\mathcal{O}_1 = I\sigma_z\sigma_z, \mathcal{O}_2 =
\sigma_zI\sigma_z, \mathcal{O}_3 = \sigma_z\sigma_zI, \mathcal{O}_4 =
\sigma_x\sigma_x\sigma_x, \mathcal{O}_5 = \sigma_x\sigma_y\sigma_y,
\mathcal{O}_6 = \sigma_y\sigma_x\sigma_y, \mathcal{O}_7
=\sigma_y\sigma_y\sigma_x$. 

Applying the linear optimal EWs on the GHZ diagonal
states given in Eq.~\ref{eq2} results in
{\begin{equation}
\fontsize{10pt}{10pt}\selectfont
\text{Tr}[\mathcal{W}\,\rho]_{_\theta}=1 \pm r_i + \cos\theta(r_j \pm r_k)+ \sin\theta(r_l \pm r_m)\label{eq4}
\end{equation}
with the values turning out to be $\geq 0$ if $\rho$ is separable, and $<0$ if
$\rho$ is entangled.

Nonlinear entanglement witnesses with nonlinear coefficients have a broader
detection range for entangled states. These nonlinear EWs are considered as an
envelope of the family of parametric linear optimal EWs
\cite{Jafarizadeh-epjd-2008} and are constructed using the envelope definition
for a family of curves \cite{weisstein-book-2002}. The family of 
nonlinear EWs are
parameterized by $\theta$ and according to the definition, the envelope of the
family of curves is obtained by simultaneously solving the one-parameter family
equations $\text{Tr}[\mathcal{W}\,\rho]=0$ and
$\frac{d}{d\theta}(\text{Tr}[\mathcal{W}\,\rho])=0$. Based on the solution of
these equations, $\theta$ for non-linear EWs is modified as
\begin{equation}
\begin{split}
    \cos\theta'&=-\frac{(r_j \pm r_k) (1 \pm r_i)}{(r_j \pm r_k)^2 + (r_l \pm r_m)^2}\\
    \sin\theta'&=-\frac{(r_l \pm r_m) (1 \pm r_i)}{(r_j \pm r_k)^2 + (r_l \pm r_m)^2}
\end{split}\label{eq7}
\end{equation}
The envelope corresponding to the non-linear EWs is given as
\begin{equation}
\fontsize{10pt}{10pt}\selectfont
    \mathcal{W}^{N}= III \pm \mathcal{O}_i + \cos\theta'(\mathcal{O}_j\pm \mathcal{O}_k)+ \sin \theta'(\mathcal{O}_l \pm \mathcal{O}_m)\label{eq5}
\end{equation}
\section{Examples of PPT entangled states}
\label{kay-key}
We have thus far discussed PPT states and explored the witnesses that are
effective for detecting entanglement in them.  From the family of available
linear and nonlinear witnesses, various combinations of Pauli operators can
effectively detect entanglement in these states. We aim to select the witness
with the highest numerical value to maximize the likelihood of experimental
observation, even in the presence of noise. This approach also applies to
nonlinear witnesses. For consistency, we use the same combination of
$(i,j,k,l,m)$ for the nonlinear witness as for the corresponding linear
witness. Since nonlinear witnesses often yield higher numerical values, they
are experimentally advantageous. We have examined families of three-qubit GHZ
diagonal PPT entangled states using both linear and non-linear EWs. We present
the chosen witnesses for entanglement detection in various bound states and the
corresponding values obtained.

\subsection{Kay three-qubit bound entangled state}
For the first example, we considered the three-qubit 
GHZ diagonal state as defined by Kay~\cite{kay-pra-2011}
\begin{equation}
\fontsize{9.5pt}{9.5pt}\selectfont
    \rho_{_{\text{Kay}}}(a)=\frac{1}{8 + 8a}
\begin{pmatrix}
4+a & 0 & 0 & 0 & 0 & 0 & 0 & 2 \\
0 & a & 0 & 0 & 0 & 0 & 2 & 0 \\
0 & 0 & a & 0 & 0 & -2 & 0 & 0 \\
0 & 0 & 0 & a & 2 & 0 & 0 & 0 \\
0 & 0 & 0 & 2 & a & 0 & 0 & 0 \\
0 & 0 & -2 & 0 & 0 & a & 0 & 0 \\
0 & 2 & 0 & 0 & 0 & 0 & a & 0 \\
2 & 0 & 0 & 0 & 0 & 0 & 0 & 4 + a \\
\end{pmatrix}
\end{equation}
The state is a valid quantum state and has positive partial transpose across
each bi-partition for $a\geq2$. It is entangled within the range of $2\leq
a\leq2 \sqrt{2}$~\cite{doherty-pra-2005}. Within this interval, the state
$\rho_{_{\text{Kay}}}(a)$ can be detected using both linear EWs as well as 
nonlinear EWs.

The state $\rho_{_{\text{Kay}}}(a)$ is detectable with 
linear and non-linear EWs of the form 
\begin{equation}
    \begin{split}
        \mathcal{W}_{_{\text{Kay}}}&=III - \mathcal{O}_1 + \cos\theta(\mathcal{O}_5- \mathcal{O}_4)+ \sin \theta(\mathcal{O}_7 - \mathcal{O}_6)\\
        \mathcal{W}^N_{_{\text{Kay}}}&=III - \mathcal{O}_1 + \cos\theta'(\mathcal{O}_5- \mathcal{O}_4)+ \sin \theta'(\mathcal{O}_7 - \mathcal{O}_6)
    \end{split}
\end{equation}
With $\theta=\pi/4$, the expectation values of the linear witness operator  Tr$[\mathcal{W}_{_{\text{Kay}}}\,\rho_{_{\text{Kay}}}(a)]$ are -0.2761,-0.09384,0 for $a=2,2.5,2\sqrt{2}$, respectively. When we choose the non-linear EW with modified $\theta'$, the expectation values of Tr$[\mathcal{W}^N_{_{\text{Kay}}}\,\rho_{_{\text{Kay}}}(a)]$ are -0.6667,-0.5714,-0.5224 for $a=2,2.5,2\sqrt{2}$, respectively.

Since the expectation values turn out to be negative in the given range, we can
conclude that PPT entanglement in the state $\rho_{\text{Kay}}$ has been
detected by the EWs. As can be seen from the above values, for a given $a$ (say
2.5), the numerical value is significant for the nonlinear case, which makes
nonlinear EWs suitable for experimental applications.

\subsection{Kye 3-qubit bound entangled state }
Next, we considered the three-qubit state defined by Kye~\cite{kye-jpa-2015}
\begin{equation}
\fontsize{9.5pt}{9.5pt}\selectfont
\rho_{_{\text{Kye}}}(b, c) = \frac{1}{6 + b + c}
\begin{pmatrix}
1 & 0 & 0 & 0 & 0 & 0 & 0 & -1 \\
0 & 1 & 0 & 0 & 0 & 0 & -1 & 0 \\
0 & 0 & 1 & 0 & 0 & 1 & 0 & 0 \\
0 & 0 & 0 & b & -1 & 0 & 0 & 0 \\
0 & 0 & 0 & -1 & c & 0 & 0 & 0 \\
0 & 0 & 1 & 0 & 0 & 1 & 0 & 0 \\
0 & -1 & 0 & 0 & 0 & 0 & 1 & 0 \\
-1 & 0 & 0 & 0 & 0 & 0 & 0 & 1 \\
\end{pmatrix} 
\end{equation}
where $b$ and $c$ are strictly positive real parameters satisfying the
condition $bc\geq 1$.  Since in the form of GHZ diagonal density matrix
(Eq.~\ref{eq1}) the diagonal elements are symmetric about the center, we choose
b=c. The state  $\rho_{_{\text{Kye}}}(b,c)$ is detectable using linear and
non-linear EWs:
\begin{equation}
    \begin{split}
        \mathcal{W}_{_{\text{Kye}}}&=III - \mathcal{O}_1 + \cos\theta(\mathcal{O}_4- \mathcal{O}_5)+ \sin \theta(\mathcal{O}_6 - \mathcal{O}_7)\\
        \mathcal{W}^N_{_{\text{Kye}}}&=III - \mathcal{O}_1 + \cos\theta'(\mathcal{O}_4- \mathcal{O}_5)+ \sin \theta'(\mathcal{O}_6 - \mathcal{O}_7)
    \end{split}
\end{equation}

We calculated the expectation values of the linear witness,
Tr$[\mathcal{W}_{_{\text{Kye}}}\,\rho_{_{\text{Kye}}}(b,c)]=-0.3314,-0.2761,-0.2367$,
for the choice of $\theta=\pi/4$ and for $b=c$ values of 2,3, and 4,
respectively. The expectation values of the non-linear witness are
Tr$[\mathcal{W}^N_{_{\text{Kye}}}\,\rho_{_{\text{Kye}}}(b,c)]=-0.4000,-0.6667,-0.8571$
for the choice of modified $\theta'$ and for $b=c$ values of 2,3, and 4,
respectively.
\subsection{Random GHZ diagonal state }
Three specific categories of states with particular parameter choices have been
identified in Ref.~\cite{Jafarizadeh-epjd-2008}, demonstrating that the
nonlinear EWs can separate the region of PPT entangled states and separable
ones completely for these special cases. This finding suggests that these
categories could be experimentally valuable. Consequently, we have considered
examples from all three categories, each based on different families of PPT
entangled states, so that we can experimentally apply both linear and nonlinear
entanglement witnesses to one state from each of these categories.\\\\

\textbf{Category 1}\\

As a first example, we have considered the state corresponding to the family
$1-r_3=r_4+r_5$ with constraints $p_4 = p_6 = p_8 = 0$, $p_3 = p_5 = p_7 = p$.
In terms of $r_k$, we get $r_1 = r_2 = r_3$ and $r_5 = r_6 = r_7$, and 
Eq.~\ref{eq2} reduces to 
\begin{equation}
\begin{split}
\rho= &\frac{1}{8}
[III + r_1 (\sigma_z\sigma_zI + \sigma_zI\sigma_z + I\sigma_z\sigma_z )+ r_4\sigma_x\sigma_x\sigma_x \\
&+ r_5(\sigma_x\sigma_y\sigma_y + \sigma_y\sigma_x\sigma_y + \sigma_y\sigma_y\sigma_x)]
\end{split}\label{eq9}
\end{equation}
This state is PPT and entangled in the region ($p_3\leq\frac{1}{4}$, $2p_1+4p_3\geq 1$,  $2p_2+4p_3\geq 1$).

We have investigated the state considering $p_1=0.2,p_2=0.35$ and
$p_3=p_5=p_7=p=0.15$. Using linear and nonlinear EWs of the form:
\begin{equation}
    \begin{split}
        \mathcal{W}&=III - \mathcal{O}_1 + \cos\theta(\mathcal{O}_4+ \mathcal{O}_5)+ \sin \theta(\mathcal{O}_6 + \mathcal{O}_7)\\
        \mathcal{W}^N&=III - \mathcal{O}_1 + \cos\theta'(\mathcal{O}_4+ \mathcal{O}_5)+ \sin \theta'(\mathcal{O}_6 + \mathcal{O}_7)
    \end{split}
\end{equation}
For this state, the expectation values of the linear entanglement witness at 
$\theta=\frac{6\pi}{5}$ and the non-linear entanglement witness are -0.2381 and -0.8, respectively.\\ 
\\
\textbf{Category 2}\\

As a second example, we consider a state  corresponding to the family
$1+r_1=r_4+r_6$. For this family, we get $p_4=0$, $p_3=p_1+p_2$ and
$p_7=p_3+p8$.\\ In this category, we investigated the state $\rho$ (Eq.
\ref{eq1}) with the choice of parameters:
$p_1=0.1,\,p_2=0,\,p_5=0,\,p_6=0.1,\,p_8=0.3$. Evaluating 
the linear witness
and the nonlinear witness:
\begin{equation}
    \begin{split}
        \mathcal{W}&=III - \mathcal{O}_1 + \cos\theta(\mathcal{O}_5- \mathcal{O}_6)+ \sin \theta(\mathcal{O}_7 - \mathcal{O}_4)\\
        \mathcal{W}^N&=III - \mathcal{O}_1 + \cos\theta'(\mathcal{O}_5- \mathcal{O}_6)+ \sin \theta'(\mathcal{O}_7 - \mathcal{O}_4)
    \end{split}
\end{equation}
we get the expectation values of linear EW at $\theta=\frac{\pi}{5}$ as -0.1587 and non-linear EW as -1.2 respectively. 
Negative expectation values of EWs show that the state is entangled for 
the given $p_k$ values.\\
\\
\textbf{Category 3}\\
\\
As a third example, we choose a state corresponding to the family $1-r_1=r_4-r_7$. For this family, in terms of $p_k$ we get, $p_1+p_3=0.5$, $p_2+p_4+p_5+p_6+p_7+p_8=0.5$.\\
In this category, we investigated the state $\rho$ (Eq. \ref{eq1}) with the choice of parameters: 
$p_1=0.2,\,p_2=0,\,p_4=0.1,\,p_5=0,\,p_6=0.2,\,p_7=0.2$. Using 
a linear witness 
and a nonlinear witness:
\begin{equation}
    \begin{split}
        \mathcal{W}&=III+ \mathcal{O}_1 + \cos\theta(\mathcal{O}_5- \mathcal{O}_6)+ \sin \theta(\mathcal{O}_7 - \mathcal{O}_4)\\
        \mathcal{W}^N&=III + \mathcal{O}_1 + \cos\theta'(\mathcal{O}_5- \mathcal{O}_6)+ \sin \theta'(\mathcal{O}_7 - \mathcal{O}_4)
    \end{split}
\end{equation}
we obtain the expectation values for linear EW at $\theta=\frac{4\pi}{15}$ and non-linear EW as -0.3298 and -0.4 respectively.
The state with these values for the parameters $p_k$, has PPT across each
bi-partition and negative expectation values of linear and nonlinear EWs,
demonstrating that the state is entangled. 
\section{Experimental Implementation}
\label{exp}
\subsection{State preparation}\label{exp1}
The IBM platform does not
permit the direct generation of mixed states. Thus, we must utilize ancilla
qubits to create a composite state involving the original state and the ancilla
qubits. By then tracing out the ancilla system, we can derive the desired mixed
state. For a three-qubit system, we would require three ancilla qubits.
Combining these ancilla qubits with the original qubits would yield a six-qubit
pure state on the IBM quantum processor. This setup allows us to effectively
examine the entanglement of the GHZ diagonal states.

\begin{figure}
\centering
\includegraphics[scale=0.89]{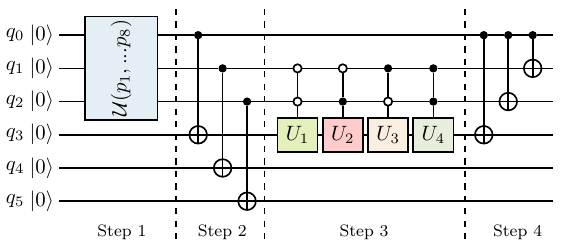}
\caption{Schematic of the 
circuit to generate a three-qubit GHZ diagonal PPT mixed state
from a six-qubit $\vert 000000 \rangle$ state. Step 1 involves applying a
unitary operation which is a function of $p_i$ values and depends on the
state to be prepared. Step 2 consists of CNOT operations. Step 3
involves another unitary operation, also a function of $p_i$ values.
Step 4 involves additional CNOT operations. }
\label{fig:protocol}
\end{figure}

We present a general protocol for creating a 6-qubit pure state 
similar to the procedure given in Ref.~\cite{hyllus-pra-2004}:
\begin{itemize}
\item {\bf Step 1}: 
Begin with an initial three-qubit 
state $\ket{000}$ and create a pure state of the form:\\
$\sqrt{\frac{p_1+p_2}{2}} (\ket{000} + \ket{100}) +\sqrt{\frac{p_3+p_4}{2}} (\ket{001} + \ket{101})$ $+\sqrt{\frac{p_5+p_6}{2}} (\ket{010} + \ket{110}) +\sqrt{\frac{p_7+p_8}{2}} (\ket{011} + \ket{111})$
\item {\bf Step 2:} Introduce three ancillary qubits to the system. Apply CNOT
gates: CNOT$_{14}$, CNOT$_{25}$, CNOT$_{36}$ (in CNOT$_{ij}$, `$i$'
denotes the control qubit, and `$j$' signifies the target
qubit).
\item {\bf Step 3}: Implement a control unitary operation 
with qubits 2 and 3 as controls and qubit 4 as the target. 
The unitary is defined as:
\begin{equation}
\begin{split}
U =& \ket{00}\bra{00} \otimes U_1 +
\ket{01}\bra{01} \otimes U_2  \\
&+
\ket{10}\bra{10}\otimes U_3+\ket{11}\bra{11} \otimes U_4
\end{split}
\label{cu}
\end{equation}

$U_1,U_2,U_3,U_4$ are given by 
\begin{equation*}
U_k= \frac{1}{\sqrt{p_i+p_{i+1}}} \left(
\begin{array}{cc}
 \sqrt{p_{i+1}}&  \sqrt{p_i} \\
  \sqrt{p_i}&  -\sqrt{p_{i+1}} \\
\end{array}\right)
\end{equation*}
where, $k=1,2,3,4$ and for $k=1 \rightarrow i=1$, $k=2 \rightarrow i=5$, $k=3 \rightarrow i=7$ and $k=4 \rightarrow i=3$.
\item {\bf Step~4}:~On applying CNOT$_{12}$,CNOT$_{13}$,CNOT$_{14}$, 
we get the desired six-qubit pure state. 
Tracing out qubits 4,5,6 will give the $\rho$ of the form in Eq.~\ref{eq1}.
\end{itemize}
This protocol enables the preparation of the desired six-qubit pure state, 
using a systematic approach for state generation, 
which involves ancillary qubits and controlled operations.

A schematic of the protocol is illustrated in Figure~\ref{fig:protocol}. The
box encompassing qubits 1,2 and 3 represents a unitary operation
$\mathcal{U}(p_1,...p_8)$ that depends on the values of $p$s.  To generate the
desired three-qubit pure state from the $\ket{000}$ state, one must determine
the values of $p_1, p_2, ...p_8$ that define the state. With this knowledge,
generating the state from the initial state becomes straightforward.  Using the
UniversalQCompiler package, unitary operations involving single and two-qubit
gates can be designed to achieve this transformation~\cite{compiler}.  The
unitaries defined in Step~2 of the protocol are standard two-qubit CNOT
operations.  To implement the general circuit given in Step~3 on the IBM
quantum processor, we decompose it into standard single-qubit gates and CNOT
gates using the UniversalQCompiler package. After applying the CNOT operations
in Step 4, the first three qubits yield the required state. The expectation
values of the witness operators are then measured on these qubits.
\subsection{Experimental Implementation on IBM }
\label{exp2}
In this study, we utilized IBM's Qiskit framework to execute and optimize
quantum computations. We used Qiskit Aer for high-performance simulation on
classical hardware, and Qiskit IBM Runtime for efficient execution on IBM's
quantum computers. Our computations were carried out on the IBM Sherbrooke
quantum processor (version 1.4.49) featuring a 127-qubit Eagle R3 processor.
These experiments aimed to explore this quantum hardware's potential for
detecting entanglement in mixed PPT states.

The IBM Sherbrooke processor uses an advanced architecture that provides higher
qubit coherence, resulting in lower gate errors despite longer gate lengths.
It has a median Echoed Cross-Resonance (ECR) gate error rate of 7.810e-3, a
median single-qubit (SX) gate error rate of 2.164e-4, and a median readout
error rate of 1.260e-2. The system offers qubit coherence times, with median T1
and T2 times of 270.58 microseconds and 193.93 microseconds, respectively.
Additional performance metrics including readout assignment error,
anharmonicity, frequency, and individual gate errors for all qubits, are
provided on the IBM website~\cite{ibm_website}.  The spatial arrangement of the
127 qubits on the IBM Sherbrooke processor is depicted in
Figure~\ref{arrangement}, illustrating the qubit connectivity and layout.

\begin{figure}[ht!]
\centering
\includegraphics[scale=0.27]{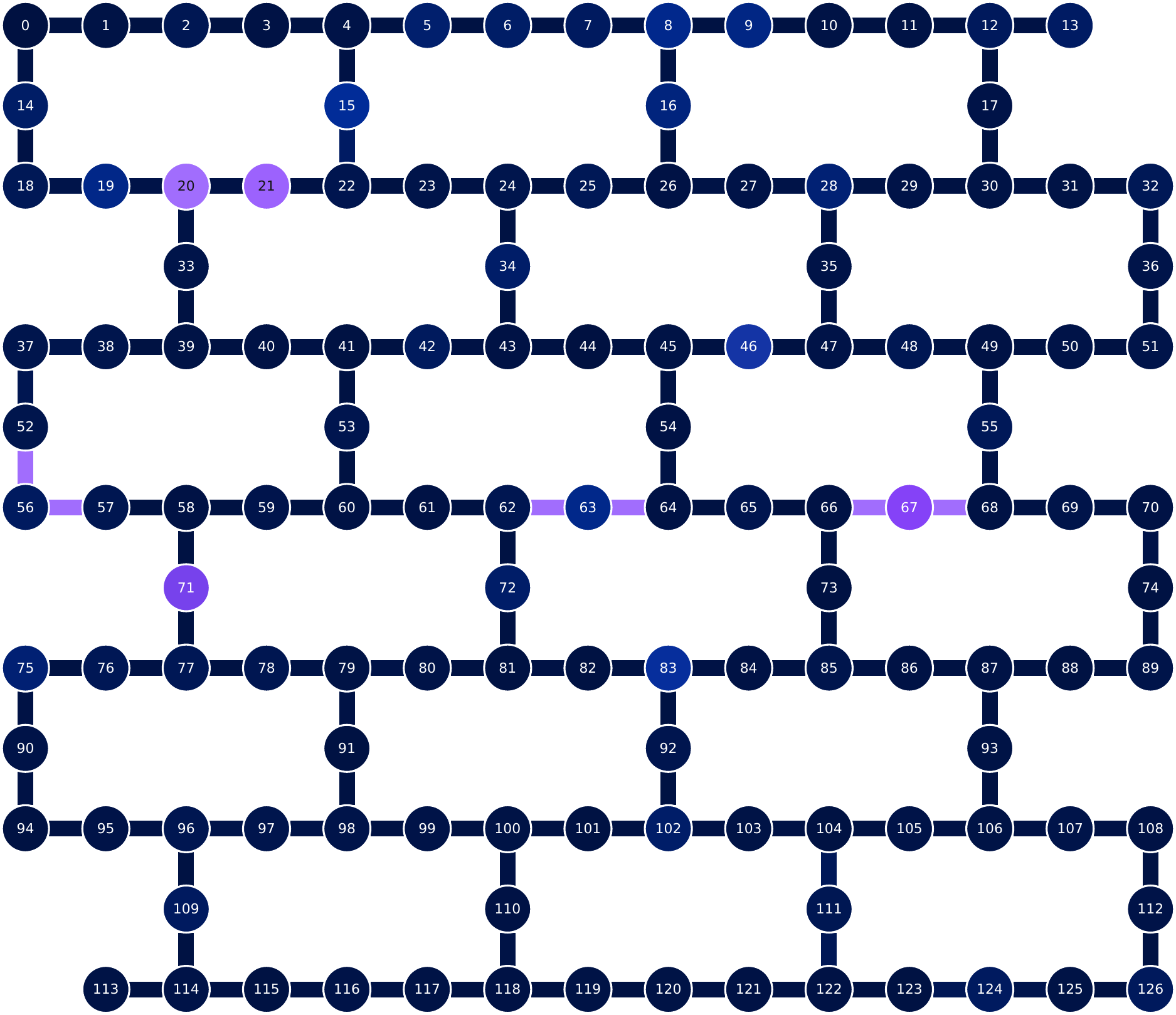}
\caption{The qubit arrangement in the IBM Sherbrooke processor 
with 127 qubits.}
\label{arrangement}
\end{figure}

We set the number of shots (the number of times the quantum circuit was
executed) to 10,000.  Before executing our quantum circuits on the IBM
hardware, we simulated all state circuits and expectation values using the
state vector simulator AerSimualtor. These simulations demonstrated excellent
agreement with theoretical predictions. This verification step ensured the
correctness of our quantum algorithms and helped identify potential issues
before running experiments on the physical quantum hardware.

\begin{figure*}[ht!]
\centering
\includegraphics[scale=0.48]{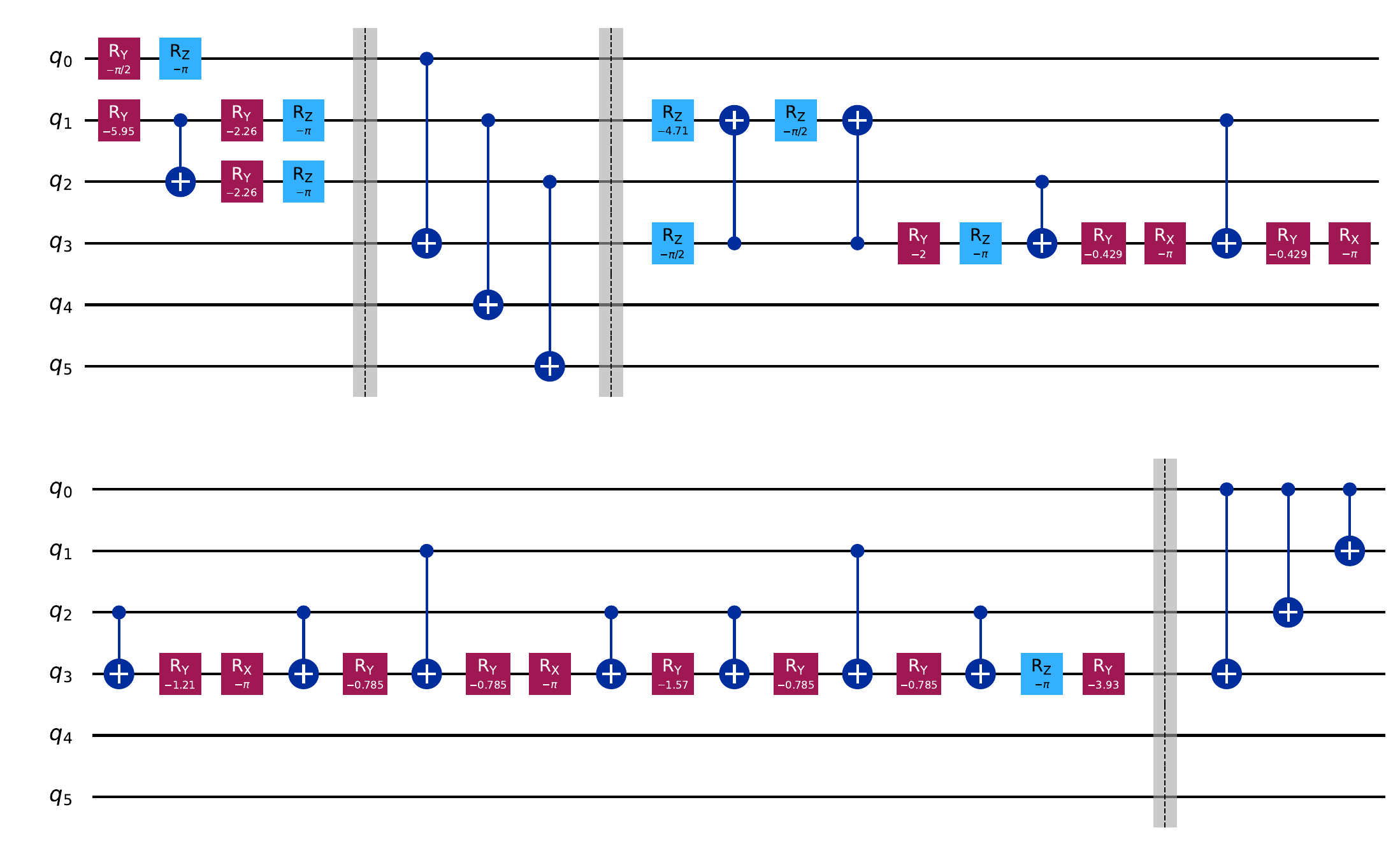}
\caption{The six-qubit gate sequence with angles and phases of the respective
single-qubit gates for the preparation of state 4 on the IBM Sherbrooke
system. The vertical lines connecting different qubits represent CNOT
gates, with dots indicating control qubits and plus symbols denoting
target qubits. Barriers are used to enhance clarity by delineating
different sections of the circuit in accordance with the various steps
of the protocol.}
\label{state1}
\end{figure*}

For our experiments, we employed the EstimatorV2 primitive in IBM Quantum
systems, part of Qiskit Runtime, which is designed to efficiently compute the
expectation values of quantum operators for parameterized quantum circuits.
Various estimator options were configured to optimize the performance of the
Estimator methods. To mitigate the effects of decoherence, we enabled dynamical
decoupling, a technique that applies a sequence of pulses to refocus the
qubits. Additionally, we utilized gate twirling and twirling of measurements to
average out errors in quantum gates and measurements, respectively. Measurement
error mitigation techniques were also applied to correct for biases in the
measurement process. 

The Estimator includes methods for error mitigation to improve the accuracy of
expectation value computations, crucial for noisy intermediate-scale quantum
(NISQ) devices. Error mitigation techniques enable users to reduce circuit
errors by modeling the device noise during execution. There are built-in error
mitigation techniques in the primitive in the form of resilience options. We
explored different resilience levels, which indicate the intensity of error
mitigation techniques applied. Resilience level 0 denotes no error mitigation,
while resilience levels 1 and 2 involve increasingly advanced techniques to
correct errors. Resilience level 2 includes techniques like Zero Noise
Extrapolation (ZNE), which estimates results as if there were no noise by
running computations at different noise levels and extrapolating back to zero
noise. By conducting experiments across these different resilience levels, we
were able to assess the effectiveness of various error mitigation strategies
and their impact on the accuracy of our quantum computations.

Transpilation is the process of rewriting a given input circuit to match the
topology of a specific quantum device, and/or to optimize the circuit for
execution on noisy quantum systems. The transpilation process was
employed to convert our high-level quantum circuits into a lower-level
representation suitable for execution on the IBM Sherbrooke processor. During
this process, optimization level 3, the highest level of optimization in
Qiskit, was applied to reduce the gate count and circuit depth, thereby
improving the fidelity of the quantum computations.

To ensure the reliability of our results, we repeated each experiment three
times for each code and each level of resilience, and then calculated the mean
and standard deviation of the results. This approach allowed us to quantify the
variability and consistency of the measurements under different conditions.

{\centering
\fontsize{9pt}{9pt}\selectfont
{\setlength{\extrarowheight}{5pt}
\begin{table*}[t!]   
\label{table1}  
\begin{tabular}{
|c|c|c|c|c|c|c|c|c| }
 \hline
&\multicolumn{2}{c|}{Theoretical}&\multicolumn{6}{c|}{Experimental Witness}\\
 \cline{4-9}
 &\multicolumn{2}{c|}{Witness}&\multicolumn{2}{c|}{Resilience=0}&\multicolumn{2}{c|}{Resilience=1}&\multicolumn{2}{c|}{Resilience=2}\\\cline{2-9}
 State & Linear &Non-Linear& Linear &Non-Linear& Linear &Non-Linear& Linear &Non-Linear\\
 \hline
1  & -0.2761 & -0.6666 & -0.0275$\pm$0.0188 & -0.5470$\pm$0.0266 & -0.0074$\pm$0.0525 & -0.5166$\pm $0.0178 & -0.2881$\pm$0.0236 & -0.6561$\pm$0.0183 \\
 2& -0.3314 & -0.4000 & -0.0580$\pm$0.0223 & -0.3068$\pm$0.0249 & -0.0371$\pm$0.0925 & -0.2977$\pm$0.0317 & -0.2124$\pm$0.0665 & -0.3491$\pm$0.0270 \\
3&-0.2480 & -0.8000 & 0.3560$\pm$0.0306 & -0.3412$\pm$0.0263& 0.0849$\pm$0.0229 & -0.6356$\pm$0.0322 & -0.1238$\pm$0.1045 & -0.6906$\pm$0.0957 \\
 4   & -0.1656 & -1.1904 & 0.1445$\pm$0.0209 & -0.8703$\pm$0.0806& 0.0674$\pm$0.0373 & -0.9669$\pm$0.0231 & -0.1495$\pm$0.0666 & -1.1916$\pm$0.0786 \\
5&   -0.3313 & -0.4000 & -0.0240$\pm$0.0234 & -0.2952$\pm$0.0120 & -0.1104$\pm$0.0323 & -0.3519$\pm$0.0025 & -0.2669$\pm$0.0809 & -0.3610$\pm$ 0.0235 \\
 \hline
\end{tabular}
\caption{Comparison of theoretical and experimental witness values for linear
and non-linear witnesses under different resilience levels.
Experimental results are shown for resilience levels 0, 1, and 2,
representing various levels of error mitigation techniques. Each
experimental value is the mean $\pm$ standard deviation of the witness
values obtained from three experimental runs.} 
\label{table1}  
\end{table*}}}

For the Kay state with $a=2$, the numerical values for the expectation values
of linear and nonlinear operators are higher. As the value of $a$ increases,
these numerical values decrease, especially for the linear witness. This
reduction in numerical values diminishes the likelihood of efficient
experimental observation. Therefore, we conduct experiments with $a=2$ to
maximize the chances of successful and efficient results.  During the
experimental setup, we designated State 1 as a Kay state with a parameter $a$
set to 2. Applying the same rationale, we opt to use the Kay state with $b=c=2$
for our experimental implementation, referring to it as State 2.  As for State
3, State 4, and State 5, they were precisely the states previously described as
examples for Category 1, Category 2, and Category 3, respectively.

To generate quantum states, gate sequences were designed based on the approach
discussed in the previous subsection, using both the algorithm and the
UniversalQCompiler package. An example of such a sequence is depicted in
Figure~\ref{state1}, where we illustrate the quantum circuit implemented for
State 4 on the IBM Sherbrooke processor.

We conducted experiments on five quantum states by designing circuits as
discussed above and using an estimator to determine expectation values, which
are essential for calculating both linear and nonlinear quantum witnesses. We
utilized estimators with three different resilience levels: 0, 1, and 2. The
resilience levels indicate the intensity of error mitigation techniques
applied, with higher levels providing more robust error correction.  For
resilience level 0 (no error mitigation) and 1 which includes basic error
mitigation techniques, the IBM Sherbrooke quantum processor took an average of
18-20 seconds to execute the code. For resilience level 2, which incorporates
advanced error mitigation techniques like Zero Noise Extrapolation (ZNE), it
took 1 minute and 6-10 seconds.  This demonstrates that while higher resilience
levels improve the accuracy of results, they also require more time to execute.
After creating and optimizing (transpiling) the circuits, we executed each code
for each state and each resilience level three times. This repetition helped us
gather sufficient data to calculate the average behavior and the standard
deviation, providing insights into the variability and consistency of the
results. The detailed experimental outcomes as well as theoretically expected
results for all states are presented in Table~\ref{table1}. Each code execution
involved 10,000 shots, ensuring robust statistical data for analysis.

It is noteworthy that while all states exhibited negative values when analyzed
using nonlinear witnesses, the linear witnesses failed to detect entanglement
in some of the states.  We observe that with resilience levels 0 and 1, the
results obtained using linear entanglement witnesses are unreliable. However,
with resilience level 2, although the results are somewhat reliable, they
require longer execution times. On the other hand, nonlinear entanglement
witnesses produce reliable results even with resilience level 0. These
experiments were conducted under maximum optimization, specifically
optimization level 3. This observation underscores the significance of
nonlinear entanglement criteria in distinguishing the entangled nature of these
states in NISQ devices.
\section{Conclusion}
\label{concl}
We examined the entanglement of three-qubit GHZ
diagonal bound states by analyzing both linear and nonlinear entanglement
witnesses on an IBM quantum processor. The states were prepared by taking three
ancilla qubits and expectation values were calculated using the Estimator
primitive on IBM runtime. We checked and compared the results of linear and
nonlinear witnesses with theoretical expectations using various error
mitigation techniques. The presence of entanglement is indicated by negative
expectation values of the entanglement witness operators. 
For all the states prepared on the
IBM quantum platform, the nonlinear expectation values were negative, even
without error mitigation, providing a strong indication of entanglement.
However, the linear expectation values for some states were positive and not
very reliable across different resilience levels of error mitigation. This
suggests that linear witnesses alone may not be sufficient for reliably
detecting entanglement in these cases. This discrepancy highlights the
importance of considering nonlinear effects in entanglement analysis. Our
findings contribute to a deeper understanding of entanglement detection in
near-term quantum computing devices.

\section*{Acknowledgment}
VG expresses gratitude to Ritajit Majumdar, \textit{IBM Quantum}, IBM India
Research Lab, for insightful discussions.  G.S. acknowledges UGC India for
financial support.

%
\end{document}